\documentclass[aps,prd,nofootinbib,preprintnumbers,11pt,
tightenlines,superscriptaddress]{revtex4}

\usepackage{graphicx}
\usepackage{rotating}
\usepackage{latexsym}

\newcommand{\beq}{\begin{equation}}
\newcommand{\eeq}{\end{equation}}
\newcommand{\be}{\begin{equation}}
\newcommand{\ee}{\end{equation}}
\newcommand{\bqa}{\begin{eqnarray}}
\newcommand{\eqa}{\end{eqnarray}}

\begin{document}

\preprint{BI-TP 2006/14}

\title{The Unstable Glasma}

\author{Paul Romatschke}
\affiliation{Fakult\"at f\"ur Physik, Universit\"at Bielefeld,
D-33501 Bielefeld, Germany}
\author{Raju Venugopalan}
\affiliation{Department of Physics, Bldg. 510 A, Brookhaven National Laboratory, Upton, NY-11973, USA}

\date{\today}

\begin{abstract}
We discuss results from 3+1-D numerical simulations of  SU(2) Yang--Mills equations for an unstable Glasma expanding into the vacuum after a 
high energy heavy ion collision. We expand on our earlier work on
a non-Abelian Weibel instability in such a system and study the behavior of the instability in greater detail on
significantly larger lattices than previously. We establish the time
scale for the onset of the instability and  demonstrate that the
growth rate is robust as one approaches the continuum limit. 
For large violations of boost invariance, non-Abelian effects cause the growth of soft modes to saturate. At late times, we observe significant 
creation of longitudinal pressure and a systematic trend towards isotropy. These time scales however are significantly larger than those 
required for early thermalization in heavy ion collisions. We discuss additional effects in the produced Glasma that may speed up thermalization. 

\end{abstract}

\maketitle

\section{Introduction}

An outstanding theoretical puzzle in high energy heavy ion collisions is to demonstrate the  thermalization of the quark-gluon matter produced 
in these collisions. Experiments at the Relativistic Heavy Ion Collider (RHIC) at Brookhaven National Laboratory (BNL) indicate that such a 
thermalized state, the Quark Gluon Plasma (QGP), has been formed in collisions of ultrarelativistic nuclear beams~\cite{whitepaper}.  Understanding 
thermalization from first principles in Quantum Chromodynamics (QCD)
is complicated by the interplay of several overlapping time scales. The collision itself, in the 
framework of the Color Glass Condensate (CGC) effective theory~\cite{CGC}, can be understood as the collision of coherent classical Yang--Mills 
fields~\cite{MV,KMW}. The typical momenta of these fields are characterized by the scale~\footnote{In the CGC framework, this scale is simply related, in leading order, to $\mu^2$, the 
color charge squared per unit area: $Q_s^2 = g^4\mu^2 N_c
\ln(g^2\mu/\Lambda)/ 2\pi$, where $\Lambda$ is an infrared cut-off.}
$Q_s$~\cite{GLR,MQ}. Because it is the only relevant time scale in the
problem, the formation time of gluons  after the collision is of order
$1/Q_s$~\cite{GLR,BlaizotMueller}. The nuclei are highly Lorentz
contracted -- the pressure gradients in the longitudinal direction after the collision are therefore enormous and the system expands outwards at nearly the speed of light. The initial space-time evolution of the produced classical fields is described by 
solutions of Yang--Mills equations with CGC initial
conditions~\cite{KMW,KV,KNV,Lappi}. The produced gluons begin to
scatter; the strength of their scattering is determined by an infrared
Debye mass $m_D$ which screens the range of their interactions. The
occupation number $f$ of produced gluons, initially of order of the
inverse strong coupling, $f\sim 1/\alpha_s$, decreases with time. It was believed initially that $2\rightarrow 2$~\cite{Mueller} 
and later $2\rightarrow 3$~\cite{Wong,BMSS} scattering processes could thermalize the system. Parametrically, however, these scattering 
processes take a long time of order $\tau_{\rm therm.}\sim \frac{1}{\alpha_s^{13/5}}\,\frac{1}{Q_s}$ in the ``bottom up" thermalization scenario~\cite{BMSS}.  Matter in this pre-equilibrium phase, where several time scales compete, has been termed a ``Glasma" to describe its transitory behavior from coherent Color Glass fields to thermalized Quark Gluon Plasma~\cite{LappiMcLerran}.   

It has been suggested for some time that instabilities, analogous to the Weibel instability~\cite{Weibel} in plasma physics, may play an important role in 
thermalization of the Glasma~\cite{Mrowczynski}. Recently, the specific mechanism in the ``Color Glass/Bottom Up"  scenario triggering the 
instability was identified~\cite{ALM} as arising from the change in sign of the Debye mass squared for anisotropic momentum distributions~\cite{RomatschkeStrickland}. One can view this, in the configuration space of the relevant fields, as the development of specific modes for which 
the effective potential is unbound from below~\cite{AL}. Detailed simulations in the Hard-Loop effective theory in 
1+1-dimensions~\cite{ALM,AL,RRS1} and in 3+1-dimensions~\cite{AMY,RRS2} have confirmed the existence of this non-Abelian Weibel instability. 
Particle-field simulations of the effects of the instability on thermalization have also been performed recently~\cite{DumitruNara,DumitruNaraStrickland}.

All these simulations consider the effect of instabilities in systems at rest. However, as discussed previously, the Glasma expands into the 
vacuum at nearly the speed of light. Recently, we presented first results on 3+1-D numerical simulations of Yang--Mills equations which demonstrated the existence of a non-Abelian Weibel instability in the expanding Glasma~\cite{PaulRaju1}. Such an instability was not seen in previous numerical 
simulations of the Glasma~\cite{KV,KNV,Lappi} which assumed strict boost invariance, and therefore, dynamics in 2+1 dimensions. Remarkably, arbitrarily weak 
violations of boost invariance trigger the non-Abelian Weibel
instability. Another striking feature of our simulations was that the 
%growth rate of the 
%instability 
unstable fields 
in the expanding Glasma 
%had the form 
grow proportional to $\exp(\Gamma \sqrt{g^2\mu \tau})$ as opposed to the usual exponential form. The former functional form was predicted by the authors of Ref.~\cite{ALM}. They deduced it simply from the fact that the scale for the growth rate in the expanding system 
is set by the Debye mass which depends on the proper time $\tau (=\sqrt{t^2-z^2})$ as  $m_D\propto 1/\sqrt{\tau}$. 

In this work, we present more details and expand on the results presented in Ref.~\cite{PaulRaju1}. In particular, we look at much larger transverse and 
longitudinal lattices and study the dependence of the growth rate on
the volume and continuum limits. Further, we extend the studies of
Ref.~\cite{PaulRaju1} to much larger violations of boost invariance
than considered previously. While kinematic violations of boost
invariance are small, dynamical small x quantum evolution effects
contribute to much larger violations of boost invariance. We consider
one particular realization of such violations of boost invariance. We
are able to follow all the stages of the evolution: 
we determine the time scale when the $\exp(\Gamma \sqrt{g^2\mu\tau})$ 
growth of unstable modes starts to dominate over the initial
spectrum, the saturation of the growth of these
modes and the subsequent 
generation of longitudinal pressure as the system evolves towards
isotropy. The time scales we obtain for this last stage are much larger than 
natural time scales for heavy ion collisions. 

Recently, a field theory formalism was developed to compute multiplicities in field theories with strong time dependent sources $j$
(with $j\sim 1/g$, where $g$ is the QCD coupling). The prototype for such a theory is the CGC~\cite{FrancoisRaju}. In this framework, the results of Refs.~\cite{KV,KNV,Lappi} can be 
understood as a leading order contribution to the inclusive multiplicities, while the computation performed here can be seen as an (approximate)
means of computing a piece of the next-to-leading order multiplicities. Other contributions are not included in this computation. (We 
will elaborate on these remarks in section V.B.) Alternately, 
the results of Ref.~\cite{FrancoisRaju} can be formulated as a
Boltzmann equation with a source term~\cite{FRS} -- the full next-to-leading order contributions will provide significant contributions towards thermalization. Our computations here must therefore be viewed as an essential 
but not exclusive piece of a complete computation of the evolution of the Glasma into the QGP.

This paper is organized as follows. In section II, we discuss the formulation of the problem of nuclear collisions in the CGC framework. In particular, we discuss how violations of boost invariance are implemented. We discuss details of the lattice formulation of the problem in the following section. In section IV, we discuss numerical results for the non-Abelian Weibel instability. In particular, we study how the Fourier modes of the longitudinal pressure 
grow as a function of proper time. We show explicitly the distribution
of unstable longitudinal (${\bf k}||{\bf k_z}$) modes and their evolution. 
We note a very
interesting behavior for the hardest momentum mode that is still
unstable, $\nu_{\rm max}$: while otherwise growing slowly, 
$\nu_{\rm max}$ increases dramatically
when the maximum amplitude of an unstable mode 
reaches a certain critical value.
A plausible interpretation of this effect is that when the amplitude
of the longitudinal fluctuations -- which roughly translates as the
amplitude of the unstable transverse magnetic field modes --
become large enough,
the Lorentz force on transverse (${\bf k}||{\bf k_\perp}$) ``particle" 
modes of the gluon fields is sufficient for them to acquire significant longitudinal
momenta. Hence the rapid increase in $\nu_{\rm max}$. This effect is enhanced 
for large boost invariance violating ``seeds", which are discussed further 
in section V. 
Because this ``Lorentz force effect'' is significant for large seeds,
we notice a rise in the longitudinal pressure, albeit the effect
becomes measurable only at late times. The increase in the longitudinal pressure is accompanied by a 
decrease in the transverse pressure; this supports our conjecture that hard transverse gluon modes display a significant change in their trajectories when the modes of the transverse magnetic fields are large enough. We observe a clear trend towards isotropy and quantify the change in the dependence of the energy density 
with proper time. Despite this trend, the time scales over which these effects occur are much too large to explain the early thermalization at RHIC. 
This may be because, as discussed previously, our numerical simulations may not include all the processes necessary for thermalization. In the final section, we summarize our results and discuss work in progress~\cite{PaulTuomasRaju} on thermalization in the Glasma.

\section{Nuclear collisions in the Color Glass Condensate}

We will provide here a brief review of nuclear collisions in the Color Glass Condensate framework. A more detailed discussion can 
be found in Ref.~\cite{CGC}.  We will first discuss the 2+1--dimensional boost invariant formulation of the problem before 
extending our discussion to the more general 3+1-D problem.

\subsection{Gluon production from classical fields}
In nuclear collisions at very high energies, the hard valence parton modes
in each of the nuclei 
act as highly Lorentz contracted, static sources of color charge for the soft wee parton, Weizs\"acker--Williams modes in the nuclei. 
By hard and soft, we mean large x or small x, where x is the longitudinal momentum fraction of partons in the colliding nuclei. Soft x modes can 
be understood as modes that are coherent across the longitudinal extent of the nucleus, or equivalently, $x \ll A^{-1/3}$. With increasing energy, the 
scale separating soft and hard modes shifts towards smaller values of x. How the sources are modified with this changing scale is quantified 
by a Wilsonian RG procedure--a discussion and relevant references can be found in Ref.~\cite{CGC}.  In a nuclear collision, the hard sources are
described by the current  
\beq
J^{\mu,a}=\delta^{\mu +} \rho_{1}^a(x_\perp) \delta(x^-)+
\delta^{\mu -} \rho_{2}^a(x_\perp) \delta(x^+)\,,
\label{current}
\eeq
where the color charge densities $\rho_{1,2}^a$ of the two nuclei are
independent sources of color charge on the light cone. Note that $x^\pm=(t\pm z)/2$. The $\delta$ functions represent the fact that Lorentz
contraction has squeezed the nuclei to infinitesimally thin sheets. The absence of a longitudinal size scale 
ensures that the gauge fields generated by these currents will be
boost-invariant, namely, independent of the space time rapidity
$\eta={\rm atanh} \frac{z}{t}$.
The gauge fields before the collision are obtained by solving the
Yang-Mills equations $D_\mu F^{\mu \nu}=j^\nu$, with
\beq
D_\mu = \partial_\mu+i g [A_\mu,.]\,, \qquad 
F_{\mu \nu}=\partial_{\mu} A_{\nu}-\partial_{\nu} A_{\mu}+i g [A_{\mu},A_{\nu}]\,,
\label{YME}
\eeq
the gauge covariant derivative and field strength tensor, respectively, in the
fundamental representation,  and $[A_\mu,.]$ denotes a commutator.

Gluon distributions are simply related to the Fourier transform 
$A_\mu(k_\perp)$  
of the solution of the Yang--Mills equations by $\langle A_\mu(k_\perp) A_\mu(k_\perp)\rangle_\rho$. The
averaging over the classical charge distributions is defined by the expression
\be
\langle O\rangle_\rho = \int d\rho_{1}d\rho_{2}\, O(\rho_1,\rho_2)
\,\exp\left( -\int d^2 x_\perp {{\rm Tr}\left[\rho_1^2(x_\perp)+\rho_2^2(x_\perp)
\right]
\over {2\mu^2}}\right) \, ,
\ee
and is performed independently for each nucleus with equal Gaussian weight $\mu^2$. Here $\mu^2\propto A^{1/3}$ fm$^{-2}$ denotes 
the color charge squared per unit area in each of the (identical)
nuclei with atomic number $A$. 
For very large nuclei, $\mu$ is much larger than the
fundamental QCD scale, $\mu^2 >> \Lambda_{\rm QCD}^2$ -- this
asymptotic condition justifies our applying weak coupling techniques
to the problem. Such a Gaussian weight is
justified~\cite{MV,Kovchegov,SangyongRaju} in the limit of $A \gg 1$
and $\alpha_s Y \ll 1$, providing a window $\ln(A^{1/3}) < Y <
A^{1/6}$ in rapidity\footnote{Here we refer to the momentum space
rapidity $Y = \frac{1}{2}\ln(\frac{E+p_z}{E-p_z})\equiv Y_{\rm beam} -
\ln(\frac{1}{x})$.} $Y$, which exists for large $A$. This window of
applicability, for large nuclei, can be extended to larger values of $Y$ by using the
Balitsky--Kovchegov equation~\cite{BalitskyKovchegov} to evolve the
sources $\rho_{1,2}$ to higher rapidities. This small x
renormalization-group evolution of the source densities preserves the
Gaussian structure of the sources -- however, it is now non-local and $\mu^2\rightarrow \mu^2(x_\perp-y_\perp)$. We do not expect this generalization to qualitatively modify the results discussed here. We will return to this discussion when we discuss violations of boost invariance.

Before the nuclei collide ($t<0$), a solution of the equations of motion is~\cite{MV}
\be
A^{\pm}=0 \,\,\,;\,\,\,
A^i= \theta_\epsilon(x^-)\theta_\epsilon(-x^+)\alpha_1^i(x_\perp)+\theta_\epsilon(x^+)\theta_\epsilon(-x^-)
\alpha_2^i(x_\perp) \, ,
\label{befsoln}
\ee
where, here and in the following, the transverse coordinates x,y have
been collectively labeled by the Latin index $i$.
The $\epsilon$ subscripts on the $\theta$-functions denote that they are smeared by an amount $\epsilon$ in the respective $x^\pm$ light cone 
directions. We require that the functions 
$\alpha_{m}^i(x_\perp)$ ($m=1,2$ denote the labels of the colliding nuclei) satisfy $F^{ij}=0$, namely, that they be pure gauge solutions of the 
equations of motion. 
%The above expression suggests that for $t<0$ the solution is simply the
%sum of two disconnected pure gauges. 
The functions $\alpha_m^i$ satisfy 
\beq
-D_i \alpha_{(m)}^i = \rho_{(m)}({\bf x_\perp}) \, .
\eeq
This equation has an analytical solution given by~\cite{JKMW,Kovchegov}
\beq
\alpha^i_{(m)}=\frac{-i}{g} e^{i \Lambda_{(m)}} 
\partial^i e^{- i \Lambda_{(m)}}, \qquad \nabla^2_\perp \Lambda_{(m)}=-g\, \rho_{(m)}\, .
\eeq
To obtain this result one has to assume path ordering in $x^\pm$ respectively for nuclei 1 and 2; we assume that 
the limit $\epsilon\rightarrow 0$ is taken at the final step. 

We now introduce the proper time $\tau=\sqrt{t^2-z^2}=\sqrt{2 x^+
x^-}$ -- the initial conditions for the evolution of 
the gauge field in the collision are formulated on the proper time surface $\tau =0$. They are obtained~\cite{KMW} by 
generalizing the previous ansatz for the gauge field to 
\bqa
A^i(x^-,x^+,x^\perp)&=&\Theta_\epsilon(x^-)\Theta_\epsilon(-x^+) 
\alpha_1^i(x_\perp)+
\Theta_\epsilon(-x^-) 
\Theta_\epsilon(x^+) \alpha_2^i(x_\perp)\nonumber\\
&&+\Theta_\epsilon(x^-)\Theta_\epsilon(x^+)
\alpha^i_3(x^-,x^+,x^\perp)\nonumber\\
A^{\pm}&=&\pm x^\pm \Theta_\epsilon(x^-)\Theta_\epsilon(x^+) 
\beta(x^-,x^+,x^\perp),
%\label{genansatz}
\eqa
where we adopt the Fock--Schwinger gauge  condition $A^\tau \equiv x^+ A^-+x^- A^+=0$. This 
gauge is an interpolation between the two light cone gauges $A^\pm$ on the $x^\pm=0$ surfaces respectively. 
The sole purpose of the ansatz for the gauge fields here is to obtain the unknown gauge fields 
$\alpha_3,\beta$ in the forward light cone from the known gauge fields $\alpha_{(1,2)}$ of the respective nuclei 
before the collision. This is achieved by invoking a physical ``matching condition" which requires that the Yang-Mills
equations $D_\mu F^{\mu \nu}=J^\nu$ be regular at $\tau=0$. The $\delta$-functions of the current in the Yang--Mills equations therefore 
have to cancel identical terms in spatial derivatives of the field strengths. 
Interestingly, it leads to the unique solution~\cite{KMW} 
\beq
\alpha_3^i(x^+,x^-,x^\perp)=\alpha^i_1(x^\perp)+\alpha^i_2(x^\perp),
\qquad 
\beta(x^+,x^-,x^\perp)=-\frac{1}{2} i\, g\,
[\alpha_{i,1}(x_\perp),\alpha_2^i( x_\perp)].
\label{matcond}
\eeq
Further, the only condition on the derivatives of the fields that
would lead to regular solutions are $\partial_\tau
\alpha|_{\tau=0}\,,\partial_\tau \alpha_\perp^i |_{\tau=0} =0$.
This completes the derivation of the initial conditions in the
boost-invariant case. 

Gyulassy and McLerran~\cite{gyulassy} argue that the 
equations of motion with the above boundary conditions remain unchanged
even when the fields $\alpha_{1,2}^i$ before the collision are smeared out in
rapidity to properly account for singular contact terms in the equations of motion. 
We concur and will discuss violations of boost invariance later. 

\subsection{Hamiltonian Chromodynamics in $\tau,\eta$ coordinates}

While the Yang--Mills equations discussed above can be solved perturbatively~\cite{KMW,KovchegovRischke,gyulassy}
in the limit $\alpha_s \mu \ll k_\perp$, it is unlikely that a simple analytical 
solution exists in general. The classical solutions
have to be determined numerically for $\tau>0$. The straightforward procedure 
would be to discretize the Yang--Mills equations. However, gauge invariance is best ensured by
solving Hamilton's equations on the lattice. In practice therefore, we need to construct the lattice Hamiltonian and obtain the corresponding 
lattice equations of motion.

The gluonic part of the QCD action in general coordinates takes the
form
\beq
S=- \frac{1}{2}\int d\tau d\eta dx_\perp \sqrt{-{\rm det} g_{\mu \nu}} {\rm Tr}%
\left[F_{\mu \nu} g^{\mu \alpha} g^{\nu \beta} F_{\alpha \beta}  + 
j_{\mu} g^{\mu \nu} A_{\nu} \right]
=\int d\tau d\eta dx_\perp {\mathcal L},
\label{action}
\eeq
where for convenience we choose to keep the gauge-covariant but not 
coordinate-covariant 
$F_{\mu\nu}$ defined by Eq.~(\ref{YME}).
For the purpose of solving the Yang-Mills equations for a heavy-ion
collision on a lattice, we shall work in the
light cone coordinates $\tau = \sqrt{2 x^+ x^-}$ and $\eta= \frac{1}{2}\ln(\frac{x^+}{x^-})$. 
Our gauge condition $x^+ A^-+x^- A^+=0$ transforms to $A^\tau=0$,
and the Lagrangian density becomes
\beq
{\mathcal L} = \tau {\rm Tr}%
\left[\frac{F_{\tau \eta}^2}{\tau^2}+F_{\tau i}^2-\frac{F_{\eta
i}^2}{\tau^2}-\frac{F_{i j}^2}{2}+\frac{j_\eta
A_\eta}{\tau^2}\right],
\eeq
due to the non-trivial metric $g_{\mu \nu}={\bf
diag}(1,-1,-1,-\tau^2)$. 
Note that the contribution of the hard valence current to ${\mathcal L}$ drops out in the boost-invariant case~\footnote{As we shall see, the 
dependence on the source densities is contained in the initial conditions through the dependence of the gauge fields on the source densities.} and can be neglected as long as one only considers a small region around central space-time rapidities $\eta=0$.
The conjugate momenta are then
\beq
E_i=\frac{\partial {\mathcal L}}{\partial (\partial_\tau A_i)}=\tau
\partial_\tau A_i\,,\qquad E_\eta=\frac{\partial {\mathcal L}}{\partial
(\partial_\tau A_\eta)}=\frac{1}{\tau}
\partial_\tau A_\eta
\eeq
which we use to construct the Hamiltonian density
\beq
{\mathcal H}=E_i(\partial_\tau A_i) +E_\eta(\partial_\tau A_\eta)-
{\mathcal L}
={\rm Tr}\ \left[\frac{E_i^2}{\tau}+\frac{F_{\eta i}^2}{\tau}+\tau
E_\eta^2 + \tau F_{x y}^2\right].
\label{contHamil}
\eeq
Using finally
\beq
\frac{\partial {\mathcal H}}{\partial E_\mu}=\partial_\tau A_\mu,\qquad
\frac{\partial {\mathcal H}}{\partial A_\mu}=-\partial_\tau E_\mu \, ,
\eeq
one finds the equations of motion in the Hamiltonian formulation,
\bqa
\partial_\tau A_i=\frac{E_i}{\tau}, &\qquad& \partial_\tau A_\eta =
\tau E_\eta \nonumber\\
\partial_\tau E_i =\tau D_j F_{ji}+\tau^{-1} D_\eta F_{\eta
i} &\qquad &
\partial_\tau E_\eta =\tau^{-1}D_j F_{j\eta}.
\eqa
The Gauss law constraint is simply
\beq
D_i E_i + D_\eta E_\eta =0 \, .
\eeq

If the sources are strictly $\delta$-function sources on the light cone, the Yang--Mills equations 
are independent of the space--time rapidity. One therefore has 
\beq
A_i(x_\perp,\eta,\tau) \equiv A_i(x_\perp,\tau)\,\,;\,\, A_\eta \equiv \Phi (x_\perp,\tau) \, ,
\ee
which results in $F_{\eta i} = -D_i\Phi$. The Hamiltonian in 
Eq.~(\ref{contHamil}) then reduces in the 
boost invariant case to 
\beq
{\mathcal H}=E_i(\partial_\tau A_i) +E_\eta(\partial_\tau A_\eta)-
{\mathcal L}
={\rm Tr}\ \left[\frac{E_i^2}{\tau}+\frac{(D_i \Phi)^2}{\tau}+\tau
E_\eta^2 + \tau F_{x y}^2\right] \,,
\label{cont-boostHamil}
\eeq
which is the QCD Hamiltonian in 2+1-dimensions coupled to an adjoint scalar field. Boost invariance clearly greatly  
simplifies the numerical simulations of these equations of motion. 

The initial conditions for the gauge-fields in the boost-invariant
case from Eq.~(\ref{matcond}) become
\beq
A^i(x_\perp)=\alpha_{(1)}^i(x_\perp)+\alpha^i_{(2)}(x_\perp), \qquad
A_\eta(x_\perp)=0, \qquad
E_i(x_\perp)=0, \qquad
E_\eta(x_\perp)=i\, g\, [\alpha_{(1)}^i,\alpha_{(2)}^i].
\label{initcond}
\eeq
The magnetic fields being defined as $B_k = \epsilon_{k\mu\nu}F^{\mu\nu}$, these initial conditions suggest that 
$B_\eta \neq 0$ and $B_i =0$. Note that the latter condition follows from the constraint on the derivatives of the gauge field that 
ensure regular solutions at $\tau=0$. Thus one has $E_\eta, B_\eta \neq 0$ and large, as well as $E_i,B_i=0$. This is in sharp contrast to 
the electric and magnetic fields of the nuclei before the collision --
they are purely transverse! Some of the consequences of these initial conditions were 
discussed previously by Kharzeev, Krasnitz and Venugopalan~\cite{KKV}; their importance was emphasized recently by Lappi and McLerran~\cite{LappiMcLerran}.

\subsection{Violations of Boost-Invariance}

In heavy-ion collisions, one clearly does not have exact boost-invariance. 
Besides simple geometric effects (a heavy nucleus can never be Lorentz contracted into an infinitely thin sheet), 
one also has to take into account quantum fluctuations at high energies.  These 
lead to violations of boost invariance that
are of order unity over rapidity scales
$Y\sim \alpha_s^{-1}$.
Consequently, a more realistic model of a heavy-ion collision will
have color sources that have a finite width in $x^\pm$
instead of the idealized $\delta$ functions of Eq.~(\ref{current}).
Nevertheless, the nuclei are highly localized on the light cone, and the same 
matching conditions required for regular solutions in the forward light cone apply. 

We will assume here that the initial conditions in Eq.~(\ref{matcond}) can be extended to the boost non-invariant case. 
Ignoring details of exact matching of the Yang--Mills equations on the light cone, the qualitative difference with respect to 
idealized boost-invariant case are rapidity fluctuations due to violations of boost invariance. In what follows, 
we will study two models of initial conditions containing rapidity fluctuations~\footnote{These models of violations of boost invariance 
are only weakly motivated at present. The only condition imposed is that these satisfy Gauss' law.  A more complete theory should  specify, from 
first principles, the initial conditions in the boost non-invariant case.}.  We construct these by
modifying the boost-invariant initial conditions  in Eq.~(\ref{initcond}) to 
\beq
E_i(x_\perp,\eta)=\delta E_i(x_\perp,\eta), \qquad
E_\eta(x_\perp,\eta)=i\, g\, [\alpha_{(1)}^i,\alpha_{(2)}^i]+
\delta E_\eta(x_\perp,\eta)
\label{initcond2}
\eeq
while keeping $A_i,A_\eta$ unchanged. The rapidity dependent
perturbations $\delta E_i,\delta E_\eta$ are in principle arbitrary, 
except for the requirement that they satisfy the previously mentioned Gauss law. For these initial conditions, it takes the 
form
\beq
D_i \delta E_i+D_\eta E_\eta=0 \, .
\eeq
We construct these perturbations as follows: 

i) we first generate random configurations $\delta \overline{E}_i(x_\perp)$ with 
$\langle\delta \overline{E}_i(x_\perp) \delta \overline{E}_j(y_\perp)\rangle =
\delta_{ij} \delta(x_\perp-y_\perp)$.

ii) Next, for our first model of rapidity perturbations, we generate a Gaussian random function $F(\eta)$ with amplitude $\Delta$
\beq
<F(\eta) F(\eta^\prime)>=\Delta^2 \delta(\eta-\eta^\prime).
\eeq
For the second model, which we shall discuss further in section \ref{sec:LS}, 
we generate the Gaussian random function but then remove the high-frequency components of $F(\eta)$ by applying 
a ``band-filter'' with strength $b$:
\beq
F(\eta)\rightarrow F(\eta)=\int \frac{d\nu}{2\pi} \exp(-i \nu \eta) 
\exp(-|\nu| b)
\int d\eta
\exp(i \eta \nu) F(\eta)\,.
\label{bandfilter}
\eeq
The white noise Gaussian fluctuations of the previous model ensure that the amplitudes of all modes are of the same size. 
The high momentum modes dominate bulk observables such as the pressure. The instability we will discuss shortly is sensitive to 
the infrared modes but its effects are obscured by the higher momentum modes. This is particularly acute for large 
violations of boost invariance. Damping the high frequency modes of the white noise spectrum will therefore allow us to also study larger values of $\Delta$, or ``large seeds'' that violate boost-invariance.

iii) For both models, once $F(\eta)$ is generated, we then obtain 
\beq
\delta E_i(x_i,\eta)=\partial_\eta F(\eta) \delta
\overline{E}_i(x_\perp), \qquad E_\eta(x_i,\eta)=-F(\eta) D_i \delta
\overline{E}_i(x_\perp) \, .
\eeq
This construction explicitly satisfies Gauss' law.

We note that in order to implement rapidity fluctuations in the above manner, 
one has to have $\tau>0$. This is a consequence of the chosen
coordinates and does not have a physical origin. We will therefore will use these initial conditions at
$\tau=\tau_{\rm init}$ with $0<\tau_{\rm init}\ll Q_s^{-1}$ and show
our results are fairly independent on
the specific choice of $\tau_{\rm init}$.

\section{Lattice Simulations}

The Hamiltonian formulation in the last section is very convenient
since it allows for a straightforward discretization that can be used
to solve the Yang-Mills equations on a space-time lattice in
$(\tau,x,y,\eta)$. We follow previous discussions of lattice equations of motion and initial 
conditions~\cite{KV,KNV}. For the reader's convenience, these are reproduced in Appendix
\ref{disc}. We work on a lattice that has periodic
boundary conditions in the transverse plane as well as in space-time
rapidity. This is justified as long as one is restricted to small
rapidity volumes $|\eta|\ll1$. 
The lattice parameters (all of which are dimensionless) are 
\begin{itemize}
\item $N_\perp$, the number of lattice sites in the transverse
direction,
\item $N_\eta$, the number of lattice sites in the longitudinal
direction,
\item $g^2 \mu \,a_\perp$,
the lattice spacing in the transverse direction,
\item $a_\eta$, the lattice spacing in the
longitudinal direction,
\item $\tau_{init}/a_\perp$, the proper time at which the 3-dimensional
simulations are begun,
\item $\delta \tau$, the time stepping size,
\item $\Delta$, the initial size of rapidity fluctuations,
\item $b$, the strength of the ``band filter" for the large seed case (see Eq.~(\ref{bandfilter})). 
\end{itemize}

Of these, only the combinations 
$g^2 \mu \,a_\perp N_\perp \equiv g^2 \mu L$ and $a_\eta N_\eta \equiv
L_\eta$ have a transparent physical meaning~\footnote{The initial size of the rapidity fluctuations $\Delta$ and the band filter parameter $b$ will 
presumably be further specified in a complete theory. For our present purposes, they will be treated as arbitrary parameters, and results presented for a large range in their values.}. Given the energy of the collision and the size of the colliding nuclei, 
one can estimate $g^2\mu$~\cite{gyulassy,KV,KNV}; since we have periodic boundary conditions, 
$L^2 = \pi R^2$. For RHIC collisions of gold nuclei, one has $g^2\mu L \approx 120$; collisions of lead nuclei 
at LHC energies will be a factor of two larger. Also, $L_\eta$ is the physical size of the region in $\eta$ being studied. 
For the purposes of this study, we will consider $L_\eta=1.6$ units of rapidity in most cases.
The continuum limit is approached by keeping $g^2\mu L$ and $L_\eta$  fixed while sending $\delta \tau\rightarrow 0,\ 
g^2 \mu a_\perp \rightarrow 0,\ a_\eta\rightarrow 0$.
For the 3-dimensional simulations, we still have to choose a value for
$\tau_{init}$, which should be such that for $\Delta=0$ we stay very
close to the result from the 2-dimensional simulations 
(for all of which $\tau_{init}=0$). Thus, we set
\beq
\tau_{init}=0.05 \,a_\perp,
\eeq
and later check how strongly the results obtained depend on this choice.

\section{The Weibel Instability}

The primary observables in simulations of classical Yang-Mills dynamics
are the components of the energy-momentum tensor~\cite{KNV},
\beq
T^{\mu \nu}=-g^{\mu \alpha} g^{\nu \beta} g^{\gamma \delta} F_{\alpha
\gamma} F_{\beta \delta}+\frac{1}{4} g^{\mu \nu} g^{\alpha \gamma} 
g^{\beta \delta} F_{\alpha \beta} F_{\gamma \delta}
\eeq
In particular,
\bqa
T^{xx}+T^{yy}&=&2\, {\rm Tr}\left[F^2_{xy}+E_\eta^2\right]\\
\tau^2 T^{\eta \eta}&=&\tau^{-2}\ {\rm Tr}\left[F_{\eta i}^2+E_i^2\right]
-{\rm Tr}\left[F_{xy}^2+E_\eta^2\right].
\label{Tmunu}
\eqa
Furthermore, the relation ${\mathcal H}=\tau T^{\tau \tau}\equiv \tau (T^{xx} + T^{yy} + \tau^2 T^{\eta\eta})$ 
(c.f. Eq.~(\ref{contHamil})) shows how the physical energy density
$T^{\tau \tau}$ is
related to the Hamiltonian density. Similarly, we introduce 
\beq
\tau P_\perp = \frac{\tau}{2} \left(T^{xx}+T^{yy}\right), \qquad
\tau P_L=\tau^3 T^{\eta \eta},
\eeq
which correspond to $\tau$ times the mean transverse and longitudinal
pressure, respectively. 

When studying the time evolution of rapidity-fluctuations, it is useful to introduce Fourier transforms of 
observables with respect to the rapidity. For example, 
\beq
\tilde{P_L}(\tau,\nu,k_\perp=0)=\int d \eta \exp(i \eta
\nu) \langle P_L(\tau,x_\perp,\eta)\rangle_\perp\,,
\label{FTdef}
\eeq
where $\langle \rangle_\perp$ denotes averaging over the transverse coordinates
$(x,y)$. Apart from $\nu=0$, this quantity would be strictly zero in
the boost-invariant ($\Delta=0$) case, while for non-vanishing
$\Delta$ and $\nu$, $\tilde{P_L}(\nu)$ has a maximum amplitude for one
specific momentum $\nu$. Using a very small but finite 
value of $\Delta$, this maximum amplitude is very much smaller than
the corresponding amplitude of a typical transverse momentum mode. The
system is very anisotropic in momentum space and consequently
prone to develop a Weibel-type plasma instability~\cite{Weibel}, as was shown previously 
in \cite{PaulRaju1,PaulRaju2}.

In order to illustrate some features of this instability, it is
convenient to excite very low frequency rapidity modes. Thus, contrary
to our earlier requirement, we have to use very large ($\Delta \eta
\simeq 100$) rapidity volumes; note that we only use this for
illustrative purposes and shall later consider much smaller volumes.

\begin{figure}
\begin{center}
\includegraphics[width=0.5\linewidth]{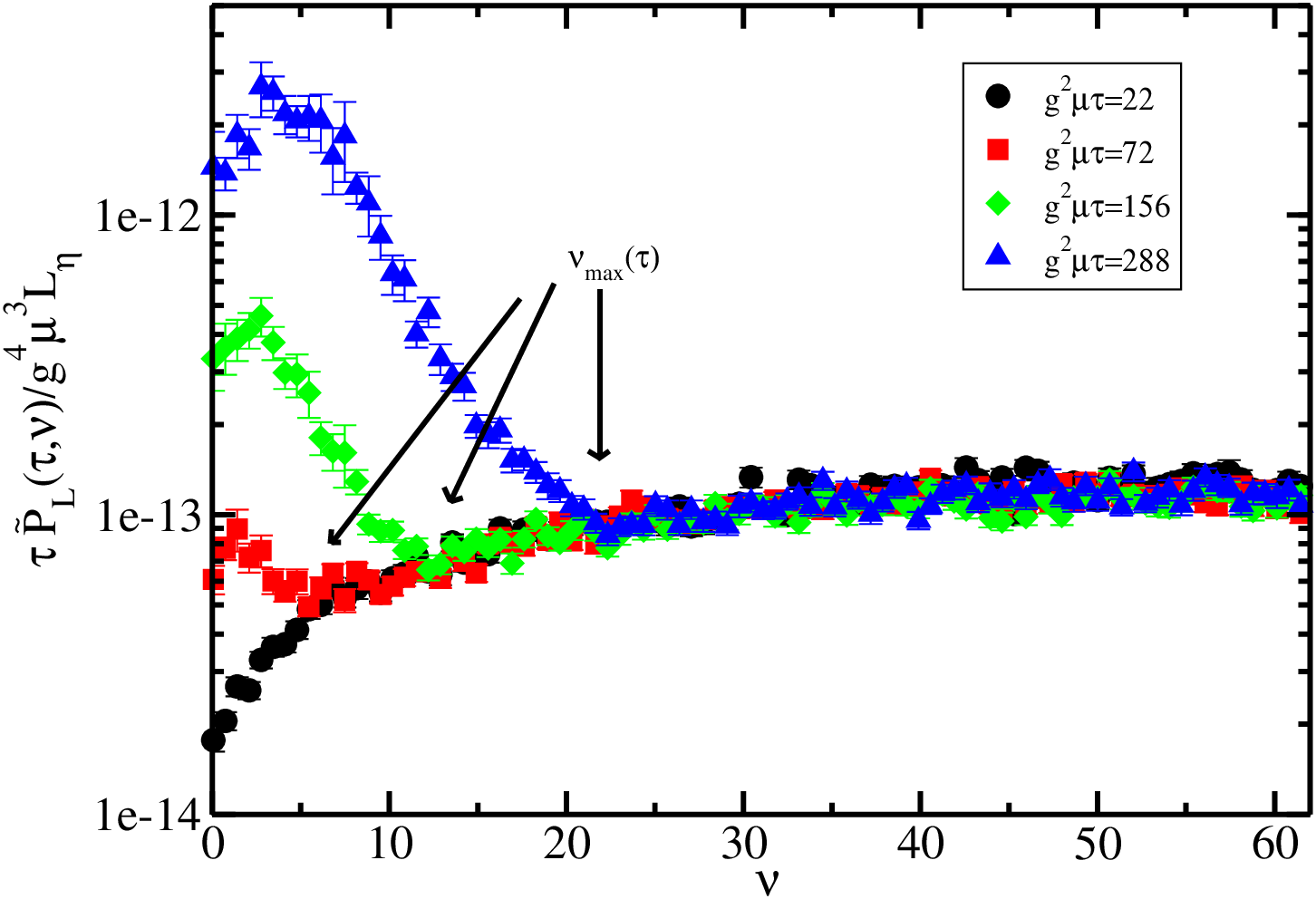}
\end{center}
\setlength{\unitlength}{1cm}
\caption{$\tau \tilde{P_L}(\tau, \nu)$ as a function of momentum
$\nu$, averaged over 160 initial conditions 
on a $16\times2048$ lattice with $g^2 \mu L=22.6$ and
$L_\eta=102.4$, $\Delta\simeq 10^{-11}$. Four different simulation
times show how the softest modes start growing with an
distribution reminiscent of results from Hard-Loop
calculations \cite{RomatschkeStrickland}. Also indicated are the respective values
of $\nu_{\rm max}$ for three values of $g^2 \mu \tau$ (see text for details).}
\label{fig:earlytimes}
\end{figure}

In Fig. \ref{fig:earlytimes} we show the ensemble-averaged 
$\tau \tilde{P_L}(\nu)$ for four different simulation times. 
The earliest time ($g^2 \mu \tau\simeq 22$) shows the configuration
before the instability sets in. Starting to overcome the effect of
expansion at times $g^2 \mu \tau\simeq
28$, the instability makes the amplitude of the soft modes grow
like $\sim \exp(\sqrt{\tau})$ (see \cite{PaulRaju1}, but also
\cite{PaulToni} for a more detailed analysis).
At times $g^2 \mu \tau\simeq 72$ the amplitudes of the softest modes
clearly differs from the starting configuration. The two later
snapshots (for times $g^2 \mu \tau\simeq 156$ and $g^2 \mu \tau\simeq
288$) indicate that the growth rate of the unstable modes closely
resembles the analytic prediction from Hard-Loop calculations~\cite{RomatschkeStrickland,ALM}. 
Fig.~\ref{fig:earlytimes} also shows that the unstable
mode with the highest mode number $\nu$ (which, in the following we will
refer to as $\nu_{\rm max}$) moves to higher $\nu$ as a function of
time. We will investigate the very interesting and suggestive behavior of $\nu_{\rm max}$ more closely in the next section.

The unstable mode with the biggest growth rate (the cusp of the
``bumps'' in Fig.\ref{fig:earlytimes}) also seems to move upwards in
$\nu$ as a function of $\tau$, but more slowly. We use this to define
the average growth rate $\Gamma^{av}$ by simply tracking the time evolution of 
the maximum amplitude of $\tau \tilde{P_L}(\nu)$ and fitting it with the
functional form
\beq
c_0+c_1 \exp{(\Gamma_{\rm fit}^{av} \sqrt{g^2 \mu \tau})}.
\label{avfit}
\eeq
The label ``average'' refers to the fact that we do not fit the growth
rate of a individual gauge mode but rather a convolution of modes
contributing to every single $\nu$ of $\tau
\tilde{P_L}(\nu)$. However, 
this method has the benefit of being a gauge-invariant
measurement, while tracking the growth of individual gauge modes
necessarily involves gauge-fixing procedures which we will address in the near future~\cite{PaulTuomasRaju}.

\subsection{Results for the growth rate}

Since the components of the energy-momentum tensor are
gauge-invariant, the fast rise of the soft longitudinal momentum
occupation numbers cannot be a gauge artifact. Instead, in principle, 
this fast rise could be due to a lattice artifact rather than a
physical instability. To convince the reader that this is not the case, we therefore present results in this 
section for the fitted average growth rates obtained for various
values of the lattice parameters. The results are shown in 
Appendix \ref{app:GR}
and indicate that the fitted average growth rate is close to 
$\Gamma_{\rm fit}^{av}\simeq 0.5$,
independent of our choices for
$a_{\eta}, g^2 \mu a_\perp, L_\eta, \tau_{\rm init}, \Delta$ and
having a weak dependence on $g^2 \mu L$. Our results suggest, unequivocally, that the 
the instability present in the Glasma is genuine and no lattice artifact.

\begin{figure}
\begin{minipage}[t]{.45\linewidth}
\begin{center}
\includegraphics[width=\linewidth]{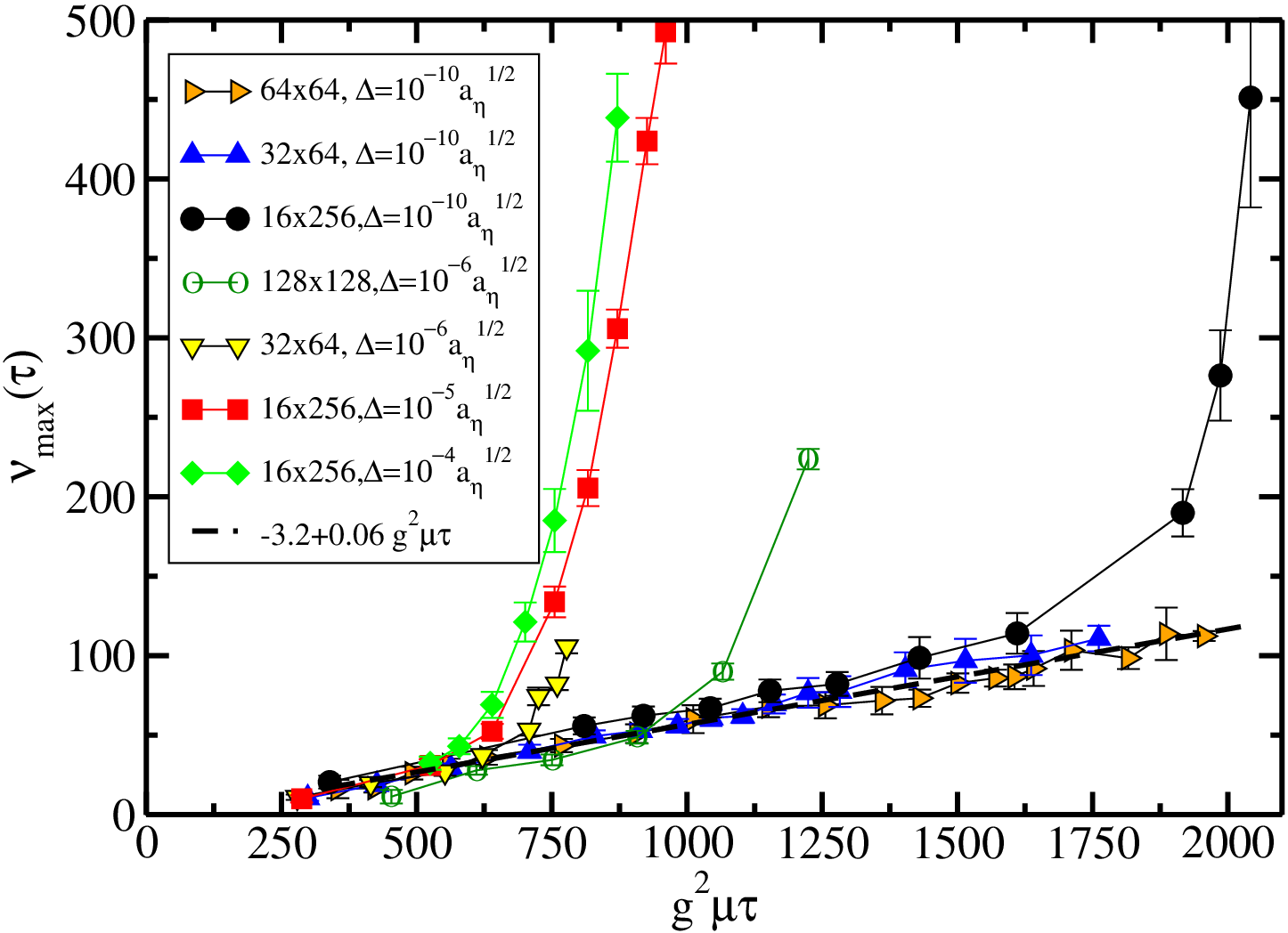}
\end{center}
\end{minipage}
\hfill
\begin{minipage}[t]{.45\linewidth}
\begin{center}
\includegraphics[width=\linewidth]{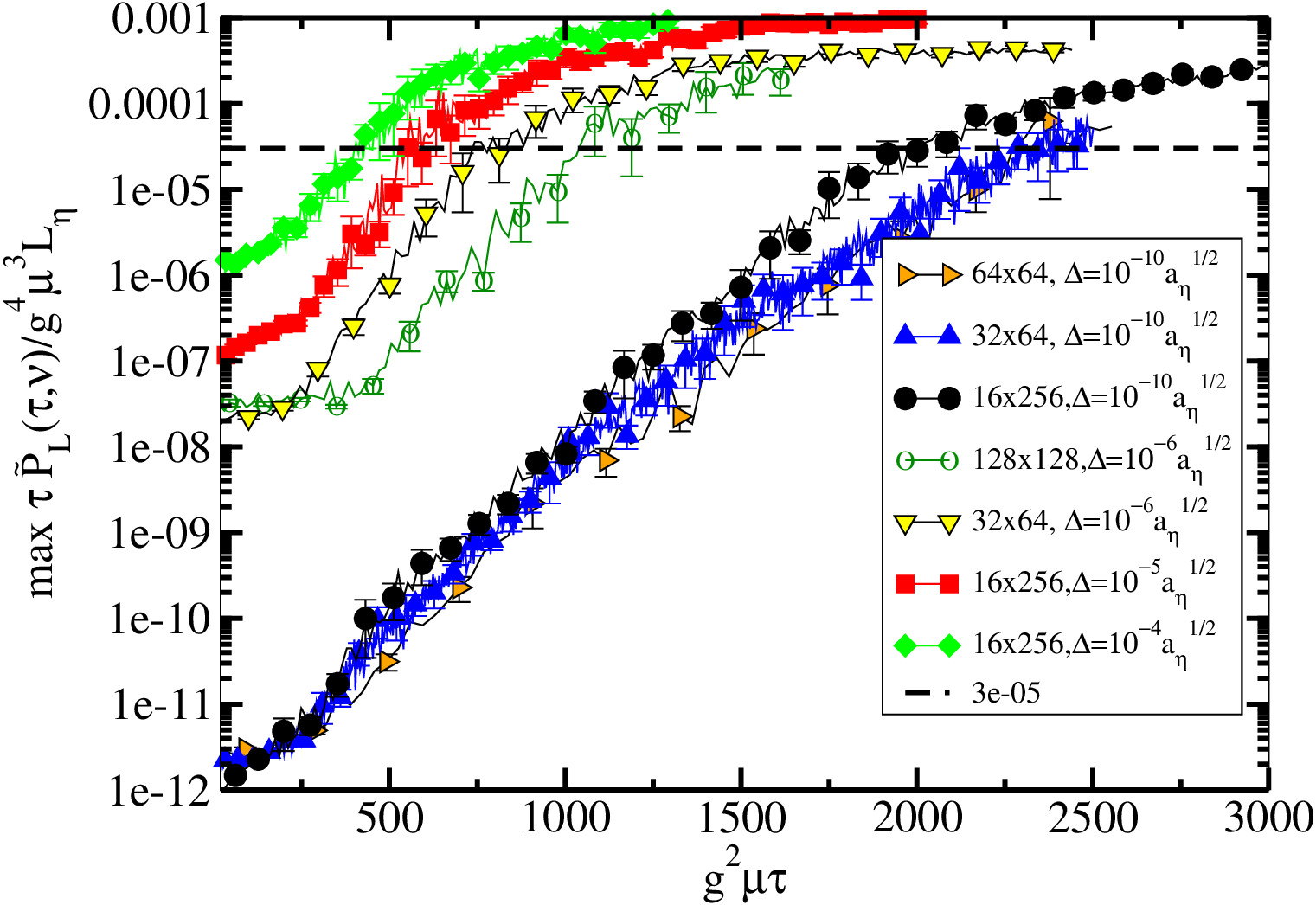}
\end{center}
\end{minipage}
\caption{Left: Time evolution of $\nu_{\rm max}$, on 
a lattices with $g^2 \mu L=22.7$, $L_\eta=1.6$ 
and various violations of boost-invariance $\Delta$. The dashed line represents the linear scaling behavior. Right: Time
evolution of the maximum amplitude $\tau \tilde P_L(\tau,\nu)$; when
this amplitude reaches a certain size (denoted by the dashed horizontal line), $\nu_{\rm max}$ starts to grow fast.}
\label{fig:numax}
\end{figure}

\subsection{Time Evolution of $\nu_{\rm max}$}

Physically, as suggested by Fig.~\ref{fig:earlytimes}, $\nu_{\rm max}$
denotes the largest mode number that is 
sensitive to the instability. The technical method we use to determine $\nu_{\rm max}$, at a given $\tau$,
is to fit a high order polynomial to the momentum spectrum 
$\tau \tilde{P_L}(\tau, \nu)$ and find
its section with a pre-defined ``baseline'' (a constant in $\nu$ and
$\tau$). Varying the ``baseline'' by a factor of $\sqrt{e}\simeq1.65$ provides
us with a rough estimate of the error of our method. 
The time evolution of $\nu_{\rm max}$ is plotted in Fig.\ref{fig:numax} for different lattices.

From this figure, one observes an underlying trend indicating 
a linear increase of $\nu_{\rm max}$ with approximately $\nu_{\rm
max}\simeq 0.06\, g^2 \mu \tau$. For sufficiently small violations of
boost-invariance, this seems to be fairly independent of the transverse or
longitudinal lattice spacing we have tested. 
 
Presumably, this can be understood as follows:  an unstable
mode with wavelength $\Delta z$ will correspond approximately to 
a wavelength $\Delta \eta$ as 
$\Delta \eta \simeq {\rm atanh}\frac{\Delta z}{t}$.
At late times (or central rapidities), we have $t\sim \tau\gg \Delta
z$ and accordingly
\beq
\nu \sim \frac{\tau}{\Delta z} \, ,
\eeq
where $\nu \sim \left(\Delta \eta\right)^{-1}$.
In other words, the wavelength $\Delta \eta$ of the unstable mode
would decrease (and $\nu$ increase) linearly as a function of $\tau$ for fixed $\Delta z$. This is
precisely what is seen in Fig.~\ref{fig:numax}.
Turning the argument around, Fig.~\ref{fig:numax} 
provides us with a measure of the unstable mode wavelength $\Delta z$
for large anisotropies,
which cannot be calculated within a Hard-Loop framework.

For much larger violations of boost-invariance -- or sufficiently late
times -- one observes that
$\nu_{\rm max}$ deviates strongly from this ``linear law''. In
Fig.~\ref{fig:numax} we show that this deviation seems to occur when
the maximum amplitude of $\tau \tilde P_L(\tau,\nu)$ reaches a
critical size, independent of other simulation parameters. This critical value is denoted by a 
dashed horizontal line and has the magnitude $3\cdot 10^{-5}$ in the dimensionless units plotted there. 
A possible explanation for this behavior is that the critical
size of the longitudinal fluctuations (corresponding to 
transverse magnetic
field modes with small $k_\perp$--see Eq.~(\ref{Tmunu},\ref{FTdef})) is sufficient 
to bend ``particle''
(hard gauge mode) trajectories out of the transverse plane into the 
longitudinal direction. This is essentially what happens in
electromagnetic plasmas. Note however that in e.m. plasmas, the particle modes are the charged fermions, while here the 
particle modes are hard ultraviolet transverse modes of the field itself.

In the next section, we will present results
that indeed show that the transverse dynamics is visibly affected once
the longitudinal fluctuations have reached a critical size. However, at the present, 
we cannot exclude the possibility that the rapid rise of $\nu_{\rm max}$ is 
due to a non-Abelian turbulent cascade as described in 
\cite{AM} or the phenomenon described in
\cite{DumitruNaraStrickland}. 
Clarifying this issue will require a 
detailed analysis of the time dependence of hard and soft modes~\cite{PaulTuomasRaju}. 

\begin{figure}
\begin{center}
\includegraphics[width=0.5\linewidth]{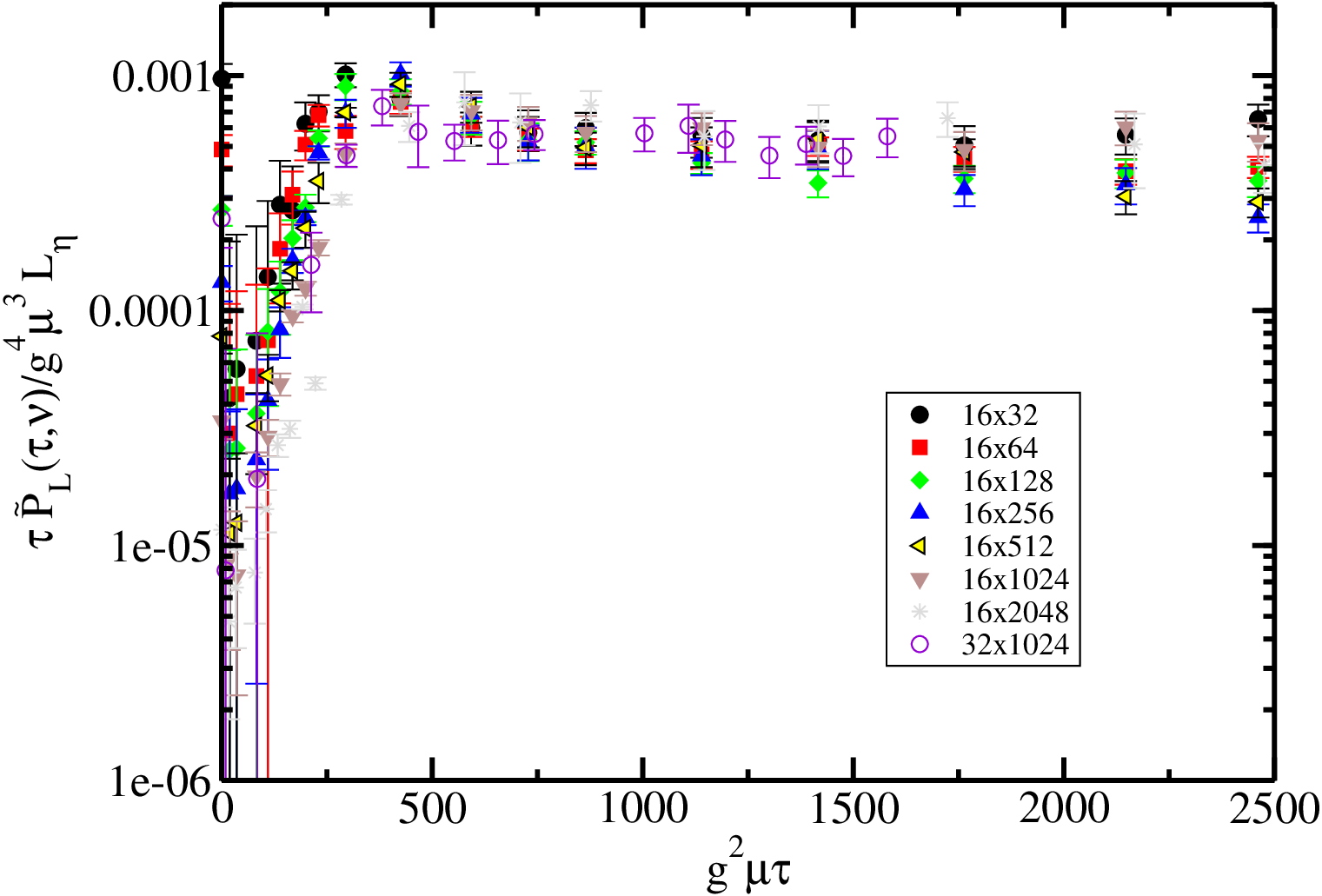}
\end{center}
\setlength{\unitlength}{1cm}
\caption{Time evolution of the (ensemble-averaged) 
maximum amplitude of $\tau \tilde{P_L}(\tau,\nu)$,
for $g^2\mu L=22.6$, $L_\eta=1.6$, $N_\perp=16,32$, $\Delta=0.1\, a_\eta^{1/2}$
and $N_\eta$ ranging from $32$ to $2048$. Larger lattices
correspond to smaller $\Delta$.  This explains why the early-time
behavior is not universal for the simulations shown here.}
\label{fig:LSsat}
\end{figure}

\section{Large Seeds}
\label{sec:LS}

In this section we focus on the (more realistic) model that involves large
and predominantly soft momentum rapidity fluctuations. The only difference
with respect to results from the previous section is the use of
Eq.~(\ref{bandfilter}), where -- unless otherwise stated -- we used the
value $b=0.5$.

In Fig.\ref{fig:LSsat}, we plot the time evolution of the maximum
amplitude of the ensemble averaged $\tau \tilde{P_L}(\tau,\nu)$, for lattices
with different $a_\eta$. Early times in this figure ($g^2 \mu \tau<200$)
correspond to the stage when the Weibel instability is operative.
Interestingly, all our simulations then show a saturation of the
growth at approximately the same amplitude. This suggests that what
one is seeing is similar to the phenomenon of ``non-Abelian
saturation'', found in the context of simulations of plasma
instabilities within the Hard-Loop framework~\cite{AMY,RRS2}.
However, while this result is robust for very large longitudinal lattices, 
further studies involving much larger transverse lattices are required to 
conclusively establish this result.

\subsection{Creation of longitudinal pressure}

\begin{figure}
\begin{center}
\includegraphics[width=0.5\linewidth]{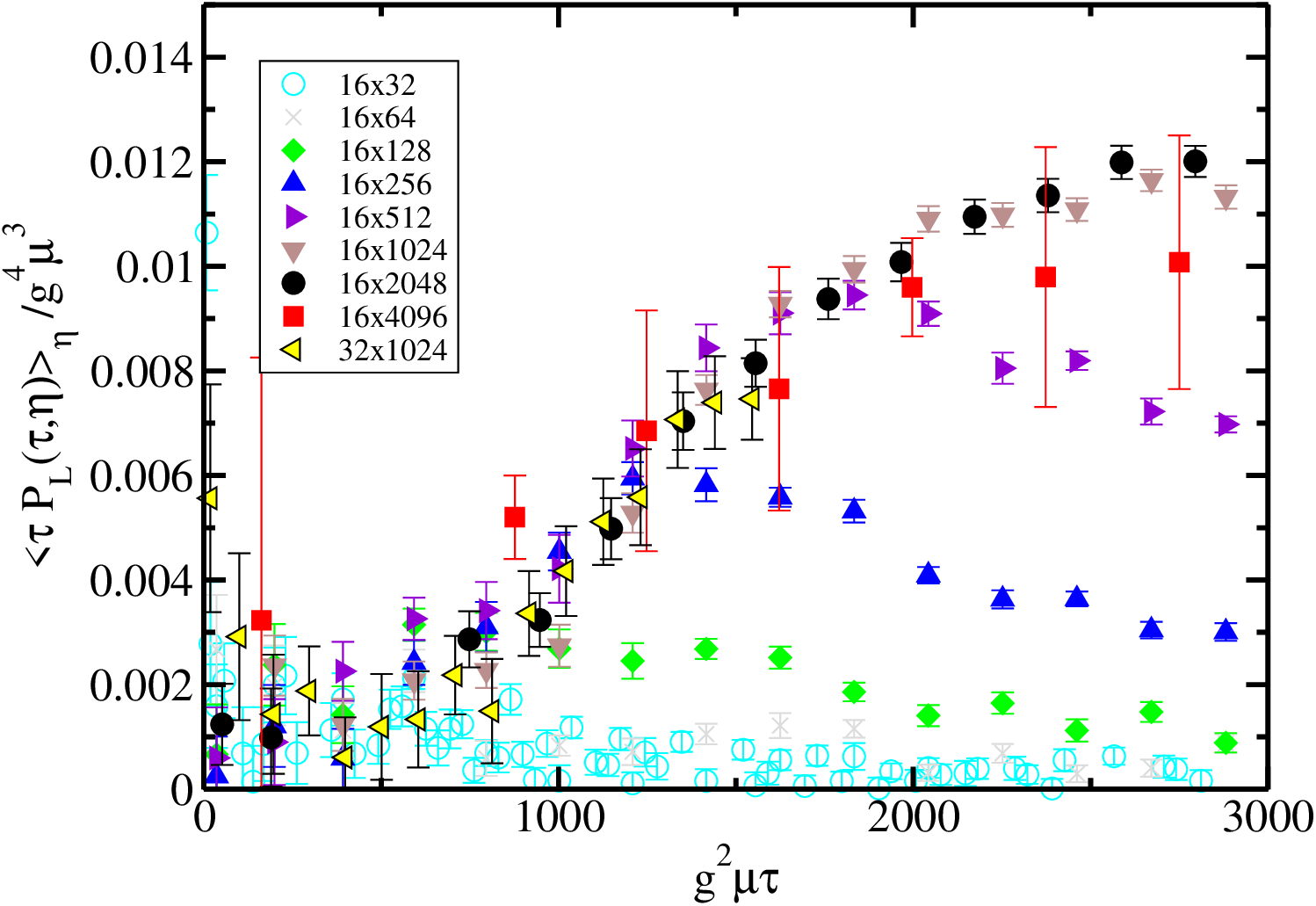}
\end{center}
\setlength{\unitlength}{1cm}
\caption{Time evolution of the ensemble and volume averaged 
longitudinal pressure $P_L$, for lattices
with $g^2\mu L=22.6$, $L_\eta=1.6$, $N_\perp=16,32$, $\Delta=0.1 a_\eta^{1/2}$
and $N_\eta$ ranging from $32$ to $2048$. A reduced statistical ensemble
of only 2 runs for $N_\eta=4096$
is consistent with $N_\eta=1024,2048$ results within error bars.}
\label{fig:Tetaeta}
\end{figure}

While for the small seed case the longitudinal fluctuations always
carried only a tiny fraction of the total system energy the situation
is different for the large seed case. In the simulations reported here, the
initial energy contained in the longitudinal fluctuations is about 1\% of the
total system energy. In this subsection, we study the behavior of the
time evolution of the average $P_L(\eta)$. This value is consistent with zero in
simulations that do not allow for longitudinal dynamics.
In Fig.~\ref{fig:Tetaeta} we plot $P_L$ as a function of $\tau$ for
different lattice spacings $a_\eta$. As can be seen, for large
$a_\eta$ (meaning low lattice UV cutoff), the longitudinal pressure is
consistent with zero while it is clearly non-vanishing when
the lattice UV cutoff is raised. However, there seems to be a limit to this rise
as there is no notable difference between the simulations for the three
smallest values of the lattice spacing.
This is suggestive that the rise in the longitudinal pressure is
physical, though again, further studies on even larger transverse lattices are
needed to strengthen this claim.

\subsection{Towards isotropy}

\begin{figure}
\begin{center}
\includegraphics[width=0.5\linewidth]{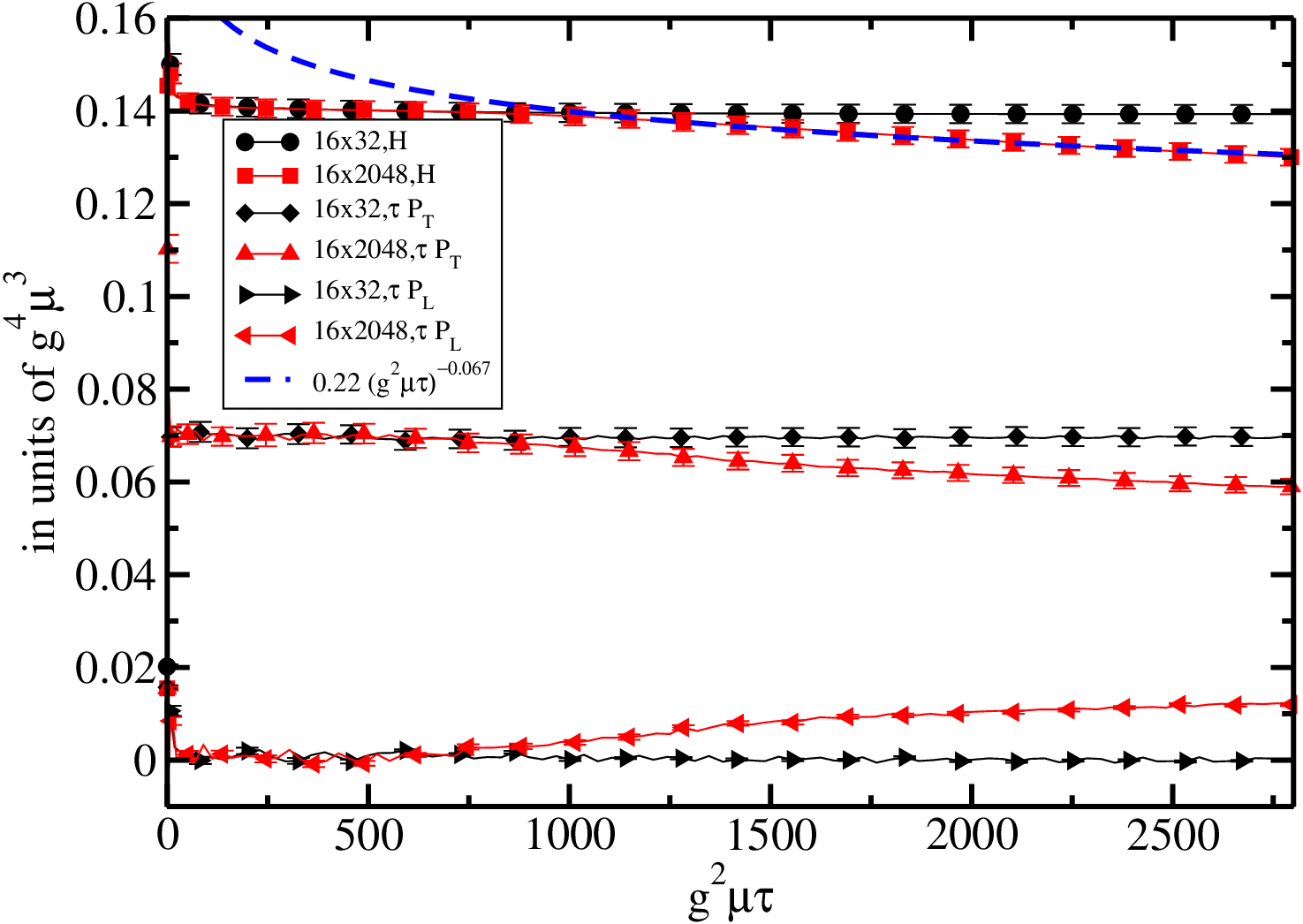}
\end{center}
\setlength{\unitlength}{1cm}
\caption{Volume and ensemble averaged Hamiltonian density $\langle H(\tau,\eta)\rangle_\eta$
(corresponding to $\tau$ times energy density) and 
transverse and longitudinal pressure 
$\tau \langle P_T(\tau,\eta)\rangle_\eta$, $\tau \langle P_L(\tau,\eta)\rangle_\eta$, for
simulations with small ($16\times32$ lattice) and large ($16\times2048$ lattice)
longitudinal UV cut-off; quantities are multiplied by $\tau$ to
facilitate comparison. The system evolves towards a
more isotropic state if the UV cut-off is sufficiently large. A fit shows that the energy density behaves as 
$\varepsilon \sim \tau^{-1.067}$ at late times. All curves calculated
on lattices with $g^2 \mu L=22.6$, $L_\eta=1.6$ and $\Delta=0.1 a_\eta^{1/2}$.}
\label{fig:isotrop}
\end{figure}

In Fig.~\ref{fig:isotrop}, we investigate the time evolution of the transverse pressure as well
as the energy density for (i) a simulation with a low UV cutoff
($16\times32$ lattice)
and (ii) a simulation with a high UV cutoff ($16\times2048$ lattice). We observe that the
rise in the mean longitudinal pressure accompanies a drop both in
the mean transverse pressure and energy density, thus bringing the system
closer to an isotropic state. The energy density depends on the proper
time as $\varepsilon \sim\frac{1}{\tau^{1.067}}$, which is distinct from 
$\varepsilon \sim \frac{1}{\tau^{4/3}}$ for an isotropic system. 
Despite this clear trend, no full isotropization was achieved in these simulations on the
time scales under investigation.  

Why is this so? We digress here to emphasize, as discussed previously, that while simulations of the classical equations of motion capture 
important aspects of the early time dynamics, they miss others. What contributions are included in the numerical simulations here, and 
what contributions are not, can be understood in a systematic formalism developed recently to treat the 
real time non-equilibrium dynamics of fields in the presence of strong ($j \sim \frac{1}{g}$) time dependent sources~\cite{FrancoisRaju}.
The power counting in these theories, as discussed in Ref.~\cite{FrancoisRaju}, is entirely in powers of $g^2 \hbar$. In this power counting, 
the 2+1-dimensional boost invariant classical solution provides the leading contribution of order $1/g^2 \hbar$ to the average 
particle multiplicity-- but includes all orders in the 
sources ($g\,j \sim O(1)$). This leading order result was computed numerically by Krasnitz, Nara and Venugopalan~\cite{KV,KNV} and by Lappi~\cite{Lappi}. At next-to-leading order (NLO), which is of order $(g^2 \hbar)^0$, there are two contributions to the 
average multiplicity: I) from the small fluctuations propagator, and II) from the product of the classical field at lowest order and the classical 
field at one loop order. NLO contributions of type I are obtained by solving partial differential equations (with retarded boundary conditions) for 
small fluctuations on top of the classical background field. The procedure followed here is very similar in spirit: we add fluctuations $\delta E_i$ 
and $\delta E_\eta$ on top of the boost invariant classical solutions and study their evolution in time. Whether these are identical to the 
NLO contributions is a topic under active investigation and beyond the
scope of this paper (but which will be addressed further in a forthcoming work~\cite{FRS}). However, it is clear that contributions similar to 
the type II NLO contributions are not included in our simulation. 

It has been argued~\cite{MuellerSon} that the early time 
classical dynamics can be smoothly matched on to a Boltzmann equation. Interestingly, the previously mentioned NLO contributions may be clearly identified in the Boltzmann equation correspondence of Ref.~\cite{MuellerSon} -- such a study would shed further light on whether the additional 
processes not included in the study here may play a role in speeding
up thermalization. 
%This discussion is beyond the scope of this work and will be addressed further in a forthcoming work~\cite{FRS}. 

In any case, since the present study is based on classical 
field dynamics only, 
reaching true thermal equilibrium is beyond the capabilities of our
simulation since description of the late stages of the evolution necessarily 
require the inclusion of quantum effects.
Nevertheless, we believe that our study is at least qualitatively valid until
the onset of the equilibration process, e.g. the point where
the energy density has departed from the free-streaming behavior
$\epsilon\sim \tau^{-1}$. As argued above, adding quantum effects through 
the full inclusion of the NLO contributions may be used to
extend the domain of applicability of such simulations.

\begin{figure}
\begin{center}
\includegraphics[width=0.5\linewidth]{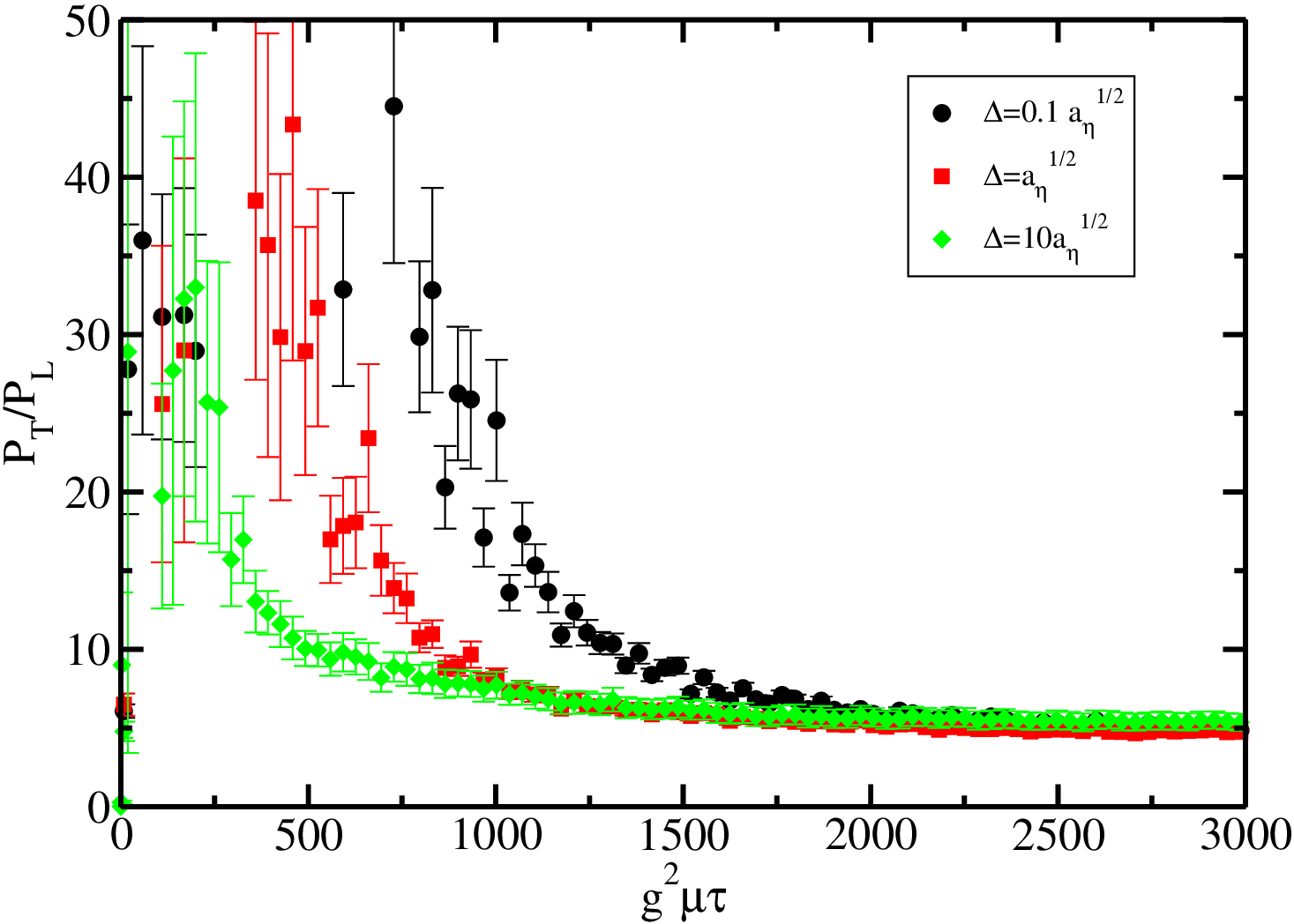}
\end{center}
\setlength{\unitlength}{1cm}
\caption{The mean system anisotropy as measured by $P_T/P_L$ for
various values of $\Delta$. Note that the late times values could be
subject to finite lattice spacing artifacts.}
\label{fig:aniso}
\end{figure}

We return to our discussion to note, however, that increasing the seed parameter $\Delta$ further
pushes the trend toward isotropization to earlier times. In Fig.~\ref{fig:aniso}, we
plot the time evolution $P_T/P_L$ as a measure of the system
anisotropy for three values of $\Delta$. One observes that larger values
of $\Delta$ accompany an earlier set-in of the isotropization
process. It should also be noted that increasing $\Delta$ further has the
effect of strongly increasing the total system energy at early times.

Similar to increasing $\Delta$, one can also decrease the value of $b$
(where we recall that this is the parameter that
controls the initial exponential fall-off of the momentum
spectrum). Decreasing $b$ by a factor of $5$ results in a system
anisotropy that is similar to the values for $\Delta=10\, a_\eta^{1/2}$
in Fig.~\ref{fig:aniso}.

We conclude here that a first principles analytic computation of $\Delta$ and $b$ would be 
extremely useful because quantitative results for the system isotropization time scale appear to 
depend on them strongly.

\section{Conclusions and outlook}

We discussed here detailed results from numerical simulations of 3+1-D SU(2) Yang--Mills equations for an 
unstable Glasma expanding into the vacuum after a high energy heavy ion collision. This work greatly expands on 
our earlier letter~\cite{PaulRaju1} where first results for 3+1-D numerical simulations were presented.
Specifically, we explained in detail the set of initial conditions we are studying, providing details of our numerical 
simulations, and tested that our results were robust in the longitudinal and transverse continuum limits. A novel addition 
to our previous studies is a detailed study of large violations of boost invariance. 

We established that for Color Glass Condensate initial
conditions, a Weibel instability is present in the system that appears  
as rapidly growing longitudinal fluctuations (starting in the far
infrared) from times of $g^2\mu \tau\sim 30$ onwards. We performed detailed checks on 
the largest available lattices that the measured
growth rate seems to be nearly independent of all our lattice
parameters. We have demonstrated that the presence of the instability is very unlikely to be
a lattice artifact.

We investigated the distribution of the unstable modes and found it to be strikingly 
similar to the analytic results from Hard-Loop calculations~\cite{RomatschkeStrickland,ALM,Schenke-etal}.
The time evolution of the maximum amplitude of these modes indicate a
saturation of the growth at a certain size.  This effect appears to be identical
to the phenomenon of non-Abelian saturation described within the
Hard-Loop framework \cite{AMY,RRS2}.
Tracking the time evolution of the hardest unstable mode, $\nu_{\rm
max}$, we deduced that the smallest unstable wavelength $\Delta z$ is
finite even for extreme anisotropies. This goes beyond results from
Hard-Loop techniques.

We find further that when the longitudinal fluctuations reach a
critical size, $\nu_{\rm max}$ increases dramatically filling up the mode spectrum. 
Because this effect is simultaneously accompanied by a drop in the transverse pressure, we interpret this result as the
effect of the Lorentz force exerted by modes of the transverse 
magnetic fields on the hard transverse gauge modes. 
These fields effectively bend the 
hard transverse gauge modes into the longitudinal direction.
Note that the total amplitude of the transverse magnetic field does
not increase significantly. Nevertheless, the ``bending''
can be accomplished since the instability has populated predominantly
modes with $k_\perp=0$ and the corresponding transverse magnetic
fields thus can act over long transverse distances.

Simulations on lattices with large longitudinal UV cutoff suggest that
significant longitudinal pressure is built up in this process.
This therefore shows a clear trend towards isotropization of the system. 
It  is also reflected in the time evolution of the energy density. The energy density
changes from the free streaming $\epsilon\sim
\tau^{-1}$ behavior to $\epsilon\sim\tau^{-1.067}$. Our results are therefore proof in 
principle that a trend towards isotropization within a weak coupling framework
is indeed possible, contrary to recent claims~\cite{Kovchegov:2005ss}. 

However, the time scales associated with isotropization in our simulations, 
are much too large in order to be of interest for phenomenology in
heavy-ion collisions. This may be  due to the fact, as discussed in section V.B,  that we are
only partly including next-to-leading order corrections in our framework~\cite{FrancoisRaju}. While these 
contributions are formally NLO and may be presumed small, they properly treat number changing processes. In this 
regard, they may be considered  leading order effects in driving the system towards equilibration. We should note further that our results strongly indicate that isotropization time scales are much shorter if longitudinal fluctuations are already large initially than 
if they have to be built up by the instability. A first principles calculation of the longitudinal gluon spectrum
after a heavy-ion collision would thus probably be a key ingredient in
understanding the process of equilibration.

\section*{Acknowledgments}
RV's research is supported by DOE Contract No. DE-AC02-98CH10886. 
PR was supported by BMBF 06BI102. We would 
like to thank  F. Gelis, T. Lappi, L. McLerran, G. Moore and M. Strickland for very useful discussions.

\begin{appendix}
\section{Lattice Discretization}
\label{disc}

Introducing the lattice gauge links $U_i=\exp(i g a_\perp A_i)$, 
$U_\eta=\exp(i g a_\eta A_\eta)$, and
making the fields dimensionless by scaling out the transverse lattice 
spacing $a_\perp$ and the coupling for convenience,
\beq
A_i\rightarrow \frac{A_i}{g a_\perp},\qquad E_i \rightarrow 
\frac{E_i}{g a_\perp}, \qquad
A_\eta\rightarrow \frac{A_\eta}{g},\qquad E_\eta \rightarrow 
\frac{E_\eta}{g a_\perp^2}
\eeq
the discretized form of the
Hamiltonian density Eq.~(\ref{contHamil}) becomes the analog of the 
Kogut-Susskind Hamiltonian,
\beq
{\cal H}_{L}=\frac{1}{\tau g^2 a_\perp^3} {\rm Tr}
\left[E_i^2+ \tau^2 E_\eta^2
+\frac{2}{a_\eta^2} \sum_i \left(1-{\rm Re} U_{\eta,i}\right)
+2 \tau^2 \sum_{\Box} \left(1-{\rm Re} U_\Box\right)
\right],
\eeq
where we also scaled the lattice time $\tau\rightarrow \tau_L =\tau/a_\perp$;
$U_\Box=U_{i,j}$ with the standard plaquette 
$U_{\mu,\nu}(x)=U_{\mu}(x)U_{\nu}(x+\mu)U_{\mu}^\dagger(x+\nu) 
U_{\nu}^\dagger(x)$.
The lattice equations of motion then become
\bqa
E_{i}^a(\tau_L+\frac{\delta \tau_L}{2},x)&=&E_{i}(\tau_L-\frac{\delta \tau_L}{2},
x)+2i\delta \tau_L\ \tau_L\ {\rm Tr}\left(
\tau^a U_{i}(\tau_L,x) \sum_{|j|\neq i}
S_{ij}^{\dagger}(\tau_L,x)\right)\nonumber \\
&&+2i\frac{\delta \tau_L}{\tau_L a_\eta^2} {\rm Tr}\left(
\tau^a U_{i}(\tau_L,x) \sum_{|\eta|\neq i}
S_{i\eta}^{\dagger}(\tau_L,x)\right)\nonumber \\
E_{\eta}^a(\tau_L+\frac{\delta \tau_L}{2},x)&=&E_{\eta}(\tau_L-\frac{\delta \tau_L}{2},
x)+2i\frac{\delta \tau_L}{\tau_L a_\eta}  {\rm Tr}\left(
\tau^a U_{\eta}(\tau_L,x) \sum_{|j|\neq \eta}
S_{\eta j}^{\dagger}(\tau_L,x)\right)\nonumber \\
U_{i}(\tau_L+\delta \tau_L,x)&=&\exp{(i \delta \tau_L\ \tau_L^{-1} \
E_{i}(\tau_L+\frac{\delta \tau_L}{2},x))} U_{i}(\tau_L,x) \nonumber \\
U_{\eta}(\tau_L+\delta \tau_L,x)&=&\exp{(i\delta \tau_L\ \tau_L a_\eta
E_{\eta}(\tau_L+\frac{\delta \tau_L}{2},x))} U_{\eta}(\tau_L,x)
\eqa
where 
\beq
S_{\mu
\nu}^{\dagger}(\tau_L,x)=U_\nu(\tau_L,x+\mu)U_\mu^{\dagger}(\tau_L,x+\nu)
U_\mu^{\dagger}(\tau_L,x)
\eeq
is the gauge link staple. For SU(2), $E^i=E^i_a \tau^a$ with
$\tau^a=\frac{1}{2} \sigma^a$ and $\sigma^a$ the
Pauli matrices and ${\rm Tr}(\tau^a \tau^b) = \frac{1}{2}\delta^{ab}$.
Note that the sum
$\sum_{|j|\neq i}$ runs over both positive and negative directions,
with $U_{-j}(\tau_L,x)=U_{j}^\dagger(x-j)$.

Numerical stability requires a very small $\delta \tau_L$ at the
beginning of the simulation, but not at late times. Thus, we have
found it convenient to use adaptive time steps, choosing
\beq
\delta \tau_L = \epsilon \frac{\tau}{\tau+T},
\label{adap}
\eeq
with $T\simeq 20$ and $\epsilon\simeq 0.025$ giving satisfactory
performance for most simulations.

\subsection{Initial conditions on the lattice}

For SU(2), the initial conditions Eq.~(\ref{initcond}) 
lead to the following form of the gauge links \cite{KV1}
\bqa
U_i&=&\left(U^{(1)}_i+U^{(2)}_i\right) \left(U^{(1)
\dagger}_i+U^{(2)\dagger}_i\right)^{-1}\\
U_\eta&=&{\bf 1}_{2}\\
U^{(1),(2)}_i(x)&=&V^{(1),(2)}(x) V^{(1),(2)}(x+i) \\
V^{(1),(2)}(x)&=&\exp(i \Lambda_{1,2}(x))\\
\Delta_L \Lambda_{1,2}(x) &=& - \tilde{\rho}_{1,2}(x)\\
<\tilde{\rho}_{1}^a(x) \tilde{\rho}_{1}^b (y)> &=& g^4 \mu^2 a_\perp^2
\delta^{ab}
\delta^2_{x,y}\\
<\tilde{\rho}_{1}^a(x) \tilde{\rho}_{2}^b (y)> &=& 0,
\label{IClatticelinks}
\eqa
where $x,y$ here denote site indices in the transverse plane and
there is no summation over $i$ on the l.h.s. of the first equation.
The last two equations should be understood such that the gauge
configurations $\rho_{1,2}$ are drawn as Gaussian random numbers 
with weight $g^2 \mu a_\perp$ while there should be no correlation
between the different nuclei. Finally, $\Delta_L$ is the lattice
Laplacian in the transverse plane,
\beq
\Delta_L \Lambda(x)=-2 \Lambda(x)+\sum_i \Lambda(x+i)+\Lambda(x-i).
\eeq
The initial condition for the momenta for the boost-invariant case are
\bqa
E_\eta^a&=&i{\rm Tr}\ \tau^a \sum_i \left(U_i^{(2)}(x)+U_i^{(1)}(x)
U^\dagger_i(x)\right.\nonumber \\
&&\left.-U^{(2)}_i(x-i)-U^\dagger_i(x-i) U_i^{(1)}(x-i)\right)-h.c.
\eqa
together with $E_i^a(x)=0$.
The rapidity fluctuations are constructed as 
\bqa
\delta E_\eta^a&=&-F(x_\eta) D_i^{a b} \delta \bar{E}_i^b (x_\perp) 
\nonumber \\
\delta E_i^a(x)&=&\left(F(x_\eta)-F(x_\eta-\eta)\right)\delta \bar{E}_i^a
(x_\perp)
\nonumber \\
<\delta \bar{E}_i^a (x_\perp) \delta \bar{E}_j^b (y_\perp)>&=&
\delta^{a b} \delta^2_{x_\perp,y_\perp} \delta_{i j}\nonumber \\
<F(x_\eta) F(x_{\eta}^\prime)>&=&a_\eta^{-1}\Delta^2 \delta_{x_{\eta} x_{\eta}^\prime}.
\eqa

For the case of large seeds, we modify $F(x_\eta)$ by creating its
Fourier-transform w.r.t. rapidity,
\beq
\tilde{F}(\kappa)=\sum_{x_\eta} \exp(2\pi i \kappa x_\eta/N_\eta)
F(x_\eta),
\eeq
removing the high-frequency modes and transforming back,
\bqa
F(x_\eta)&=&\frac{1}{N_\eta} \sum_{\kappa\leq N_\eta/2}
\exp(-2\pi i\kappa x_\eta /N_\eta) \exp(-2 \pi b |\kappa|/N_\eta/a_\eta)
\tilde{F}(\kappa)\nonumber\\
&&+\frac{1}{N_\eta} \sum_{\kappa> N_\eta/2}
\exp(-2\pi i\kappa x_\eta /N_\eta) 
\exp(2 \pi b (|\kappa|-N_\eta)/N_\eta/a_\eta)
\tilde{F}(\kappa)\nonumber.
\eqa

\section{Measured Growth Rates}
\label{app:GR}

We present here a collection of tables with measured growth rates for
various different lattice parameters.

%\vfill
\begin{minipage}{.45\linewidth}
%\begin{table}[h]
\begin{center}
\begin{tabular}{|c|c|}
\hline
$N_\eta$ & $\Gamma_{\rm fit}^{av}$\\
\hline
32  & $0.50\pm 0.01$\\%   & $0.026\pm0.001$\\
64  & $0.52\pm 0.02$\\%   & $0.030\pm0.0015$\\
128 & $0.52\pm 0.02$\\%   & $0.027\pm0.001$\\
256 & $0.524\pm 0.016$\\% & $0.031\pm0.0015$\\
& \\
\hline
\end{tabular}
\end{center}
TABLE I: Testing for the rapidity lattice spacing dependence on lattices with
$g^2\mu L=22.6, L_\eta=1.6$ with $N_\perp=16$ and 
$\Delta=10^{-10} a_\eta^{1/2}$.
%\label{tab:r1}
%\end{table}
\end{minipage}
\hfill
\begin{minipage}{.45\linewidth}
%\begin{table}[h]
\begin{center}
\begin{tabular}{|c|c|}
\hline
$N_\eta$ & $\Gamma_{\rm fit}^{av}$\\%  & $\Gamma_{\rm fit}^{ind}$\\
\hline
32 & $0.5\pm 0.01$  \\   
64 & $0.508\pm 0.012$  \\  
128 & $0.488\pm 0.01$   \\
256 & $0.492\pm 0.016$ \\
512 & $0.543\pm 0.016$ \\
\hline
\end{tabular}
\end{center}
%\caption{
TABLE II: Testing for the rapidity volume dependence on lattices with
$g^2\mu L=22.6, a_\eta=0.05$ with $N_\perp=16$ and 
$\Delta=10^{-10}a_\eta^{1/2}$.
\label{tab:v1}
%\end{table}
\end{minipage}
%\vfill

%\newpage
\vspace*{1cm}
\begin{minipage}{.45\linewidth}
%\begin{table}[h]
\begin{center}
\begin{tabular}{|c|c|c|}
\hline
$N_\perp$ & $N_\eta$ & $\Gamma_{\rm fit}^{av}$\\% & $\Gamma_{\rm fit}^{ind}$\\
\hline
16& 32 & $0.50\pm 0.01$\\
32& 64 & $0.504\pm 0.01$\\
64& 64 & $0.488\pm 0.02$\\
&&\\
\hline
\end{tabular}
\end{center}
%\caption{
TABLE III: 
Testing for the transverse lattice spacing dependence on lattices with
$g^2\mu L=22.6$,$L_\eta=1.6$ and $\Delta=10^{-10}a_\eta^{1/2}$.
%}
%\label{tab:r2}
%\end{table}
\end{minipage}
\hfill
\begin{minipage}{.45\linewidth}
%\begin{table}[h]
\begin{center}
\begin{tabular}{|c|c|c|c|}
\hline
$N_\perp$ & $N_\eta$ & $g^2\mu L$ & $\Gamma_{\rm fit}^{av}$ \\%& $\Gamma_{\rm fit}^{ind}$\\
\hline
16 & 32 & 22.6 & $0.50\pm 0.01$ \\
64 & 64 & 67.9& $0.426\pm 0.01$ \\
128 & 64 & 90.5& $0.45\pm 0.03$ \\
128 & 256 & 181 & $0.394\pm 0.04$ \\
\hline
\end{tabular}
\end{center}
%\caption{
TABLE IV: Testing for the transverse volume dependence on lattices with
$L_\eta=1.6$ and $\Delta=10^{-10}a_\eta^{1/2}$.
%\label{tab:v2}
%\end{table}
\end{minipage}

%\newpage
\vspace*{2cm}
%\begin{table}[h]
\begin{minipage}{.45\linewidth}
\begin{center}
\begin{tabular}{|c|c|}
\hline
$\tau_{\rm init}/a_\perp$ & $\Gamma_{\rm fit}^{av}$ \\%& $\Gamma_{\rm fit}^{ind}$\\
\hline
0.025 & $0.488\pm 0.014$\\
0.05 & $0.504\pm 0.01$\\
0.1 & $0.51\pm 0.01$\\
\hline
\end{tabular}
\end{center}
TABLE V: Testing for the dependence on $\tau_{\rm init}$ on 
lattices with $g^2\mu L=22.7, L_\eta=1.6$, $N_\perp=32,N_\eta=64$ and 
$\Delta=10^{-10} a_\eta^{1/2}$.
\end{minipage}
\hfill
\begin{minipage}{.45\linewidth}
%\begin{table}[h]
\begin{center}
\begin{tabular}{|c|c|c|}
\hline
$-\log_{10}{\left(\Delta a_\eta^{-1/2}\right)}$ & $\Gamma_{\rm
fit}^{av}$\\% & $\Gamma_{\rm fit}^{ind}$\\
\hline
10 & $0.504\pm 0.01$\\
6 & $0.514\pm 0.03$\\
3 & $0.48\pm 0.05$\\
\hline
\end{tabular}
\end{center}
TABLE VI: Testing for the dependence on the initial seed amplitude $\Delta$ on 
lattices with $g^2\mu L=22.7, L_\eta=1.6$, $N_\perp=32,N_\eta=64$.
\end{minipage}

%\newpage

\vspace*{1cm}
%\newpage
\begin{minipage}{\linewidth}
\begin{center}
\begin{tabular}{|c|c|c|}
\hline
$\epsilon$ & $\Gamma_{\rm fit}^{av}$\\% & $\Gamma_{\rm fit}^{ind}$\\
\hline
0.025 & $0.50\pm 0.01$\\
0.0125 & $0.50\pm 0.01$\\
0.00625 & $0.51\pm 0.01$\\
\hline
\end{tabular}
\end{center}
TABLE VII:
Testing for the dependence on the time step $\delta \tau$
lattices with $g^2\mu L=22.7, L_\eta=1.6$, $N_\perp=16,N_\eta=32$.
Note that adaptive step sizes are used (see Eq.~(\ref{adap})).
\end{minipage}
%\vspace*{1cm}

\end{appendix}

\end{document}